\documentstyle{article} 

\begin{document}

\setcounter{secnumdepth}{2}

\renewcommand{\refname}{Bibliography}

\newcommand{\refEq}[1]{(\ref{#1})}
\newcommand{\scalarproduct}[2]{\mbox{$\langle  #1 \! \mid #2 \rangle $}} 

\newtheorem{prop}{Proposition}
\newtheorem{Th}[prop]{Theorem}  
\newtheorem{lem}[prop]{Lemma}
\newtheorem{rem}[prop]{Remark}
\newtheorem{cor}[prop]{Corollary}

\title{{\bf The Darboux-Bianchi-B\"acklund transformation and soliton
surfaces}}

\author{Jan Cie\'sli\'nski
\\ {\footnotesize Uniwersytet w Bia\l ymstoku, Instytut Fizyki}
\\ {\footnotesize 
   15-424 Bia\l ystok, ul.\ Lipowa 41, Poland}
\\ {\footnotesize e-mail: \tt janek\,@\,alpha.uwb.edu.pl } }
\date{ \it Published in: {  Proceedings of First Non-Orthodox School on} {\it Nonlinearity and Geometry}, pp.\ 81-107;  edited by D.\ W\'ojcik and J.\  Cie\'sli\'nski, PWN, Warsaw 1998}

\maketitle 

\begin{abstract} 

In the first part of the paper we present the dressing method which generates
multi-soliton solutions to integrable systems of nonlinear partial differential
equations.  We compare the approach of Neugebauer with that of Zakharov, Shabat
and Mikhailov. In both cases we discuss the group reductions and reductions
defined by some multilinear constraints on  matrices of the linear problem.
The second part of the paper describes the soliton surfaces approach. The so
called Sym-Tafel formula simplifies the explicit reconstruction of the surface
from the knowledge of its fundamental forms, unifies various integrable
nonlinearities and enables one to apply powerful methods of the theory of
solitons to geometrical problems.  The Darboux-Bianchi-B\"acklund
transformation (i.e., the dressing method on the level of soliton surfaces)
reconstructs explicitly  many classical transformations of XIX century.  We
present examples of interesting classes of surfaces obtained from spectral
problems. In particular, we consider spectral problems in Clifford algebras
associated with orthogonal coordinates.
Finally, compact formulas for multi-soliton surfaces are discussed and
applied for  the Localized Induction Equation with axial flow.

\end{abstract} 

\section{Introduction}

An integrable system can be represented as integrability
conditions for a {\bf linear problem} (a system of linear partial
differential equations containing the spectral
parameter). In this paper we confine ourselves to 
linear problems of the form 
\begin{equation} \label{ZS}
\Phi,_k = U_k \Phi \ ,
\end{equation} \par \noindent
where  $x^1,x^2$  are independent variables, $\Phi,_k := \partial \Phi/\partial x^k$
and  $U_1, U_2$ are complex $n\!\times\! n$ matrices
which depend in the prescribed way on the dependent variables (shortly,
``soliton fields''),  also through their
derivatives or integrals, and on the so called {\bf spectral parameter} $\lambda$.
In practice $U_k$ are meromorphic (usually rational) functions of $\lambda$.

Considering the system \refEq{ZS} as equations for an unknown function $\Phi$
one can easily see that a non-trivial solution exists if and only if the 
following integrability condition  holds:
\begin{equation} \label{cc}
 U_1,_2 - U_2,_1 + [U_1, U_2] = 0 \ .
\end{equation} \par \noindent
The ``zero curvature condition'' \refEq{cc}, considered as an identity with repect
to $\lambda$, is equivalent to a system of nonlinear partial differential equations
for soliton fields. The system \refEq{cc} has many interesting properties which
justify to call it {\bf integrable} \cite{AS,Kri,MS,ZMNP}.
One of the properites, the existence of the Darboux-B\"acklund transformation,
is discussed thoroughly in this paper.

  The system \refEq{ZS} has $n$ independent vector solutions. It is convenient to
consider them as columns of a non-degenerate $n\!\times\! n$ matrix called the fundamental solution
of \refEq{ZS}.  In the sequel we always assume that $\Phi$, sometimes called the
{\bf wave function}, is the matrix.

\begin{rem} \label{C}
  A linear combination of vector solutions is a solution as well. In other
words, if $\Phi$ and $\Phi'$ are fundamental solutions of \refEq{ZS} then 
there exists a constant non-degenerate matrix $C$ such that $\Phi'=\Phi C$.
\end{rem}

As an example consider the {\bf nonlinear Schr\"odinger equation}:
\begin{equation}   \label{NSE}
     iq,_2 + fq,_{11} + 2q\mbox{$\mid q \mid$}^2  = 0 \ ,   
\end{equation} \par \noindent
where  $q=q(x^1,x^2)\in {\bf C}$  is the soliton field. The associated linear
problem reads \cite{AS,ZMNP},
\begin{equation}  \label{NSE-problin} 
     \Phi,_1 = (i\lambda \sigma_3 + Q) \Phi \ ,  \qquad \quad
     \Phi,_2 = ( -2i\lambda^2 \sigma_3 -2 \lambda Q + R) \Phi \ ,
\end{equation} \par \noindent	  
where $\sigma_3 := \left( \begin{array}{cc} 1 & 0 \\ 0 & -1 \end{array} \right) \ , \quad 
        Q:=\left( \begin{array}{cc} 0 & q \\ -\overline{q} & 0 \end{array} \right) \ , \quad 
		  R:=\left( \begin{array}{cc} i\mbox{$\mid q \mid$}^2 & iq,_1 \\ i\overline{q},_1 & 
		  -i\mbox{$\mid q \mid$}^2 \end{array} \right)$.

\par \vspace{1ex} \par

  The {\bf Darboux-Bianchi-B\"acklund (DBB)} transformation  is a gauge
transformation which preserves the form of the
linear problem. On the level of soliton fields this transformation simply
adds a soliton solution.
Usually the name of Bianchi is not mentioned in this context.
However, because of his enormous contribution to this field it seems justified 
to name after him at least the corresponding transformation on the level of
{\bf soliton surfaces} (see Section~\ref{SYM}).

There are several approaches to the construction of multi-soliton
solutions by gauge transformations.
Let us mention the Za\-k\-ha\-rov-Shabat dressing method \cite{LRB,ZMNP,ZS}, 
the Neugebauer approach \cite{MNS,NK}, Darboux transformation for linear
differential operators
\cite{Ma,MS}, the method of Its based on
axiomatics of the wave functions \cite{Its} 
 and the approach of Gu \cite{Gu}.
The methods have been discovered to some extent 
independently. The fundamental concept of the ``dressing''
of linear operators proposed by Zakharov and Shabat \cite{ZS1,ZS}
 was probably the starting point of all these researches.
 The details however are different and very often authors do not seem
to be aware of the results of the other groups. In this paper we
 focus on the Neugebauer and Zakharov-Shabat methods.

The dressing method of Zakharov and Shabat \cite{ZMNP,ZM-ZETF,ZS1,ZS} is a
general scheme to solve integrable nonlinear equations.  
Consider  a gauge transformation 
\begin{equation}    \label{DBTPhi}
     \tilde{\Phi} = D \Phi \ ,                          
\end{equation} \par \noindent
defined by a matrix $D=D(x^1,x^2;\lambda)$. Then, obviously, $\tilde{\Phi}$ satisfies
a linear system \refEq{ZS} with matrices $\tilde{U}_k$
\begin{equation}      \label{DBTU}
     \tilde{U}_k = D ,_k D ^{-1} + D U_k D ^{-1} \ ,  \qquad  k=1,2 \ .
\end{equation} \par \noindent
If we are able to construct a
matrix $D$ such that $\tilde{U}_k$ given by \refEq{DBTU} are of exactly the same
form as $U_k$ then $D$ is called the {\bf Darboux matrix} \cite{Ci-dbt,LRS,MS}.
In other words, the dependence of $\tilde{U}_k$ on $\lambda$
 and on soliton fields should be exactly the
same as that of $U_k$ (of course, the soliton fields entering $\tilde{U}_k$ are
in general different than those entering $U_k$).
The existence of the Darboux matrix is one of the criterions of integrability.

It turns out that the structure of matrices $U_k$ can be characterized 
completely in terms of some simple properties.
The most important information is contained in the singularities 
of $U_k$. The matrices $U_k$ are assumed to be meromorphic in 
$\lambda$ and their poles are given.

In this paper we show that the construction of
the Darboux matrix 
is practically algorithmic (see also \cite{Ci-dbt}).
The crucial point is to notice all relevant algebraic
and group properties of the associated linear problem (especially to identify
the reduction group) and then  to apply appropriate general
theorems. In particular,
we present multilinear constraints which are usually invariant with respect to
DBB transformation.  Taking them into account we can avoid some cumbersome
calculations and our construction assumes a more elegant form.

It is reasonable to consider as {\bf equivalent} the Darboux matrices which
generate the same DBB transformation on the level of soliton fields.
In other words, $D_1$ and $D_2$ are equivalent iff the coresponding
transformations \refEq{DBTU} are identical.

\begin{rem}  \label{equivalent}
 The Darboux matrix $D$ is equivalent to $f D$ where $f$ is any scalar 
 function independent on $x^1,x^2$, i.e. $f=f(\lambda)$.
\end{rem}

The methods
of constructing the Darboux matrix are based on various approaches
to matrices with rational coefficients.
The following example shows that problems of that kind can be
interesting even in the well known scalar case. Namely, 
we proceed to the decomposition into partial fractions.
The procedure (used in the integration of rational functions) is standard and
algorithmic but usually is considered as cumbersome and boring.
However, using local parameters in the neighbourhood of singularities
one obtains a very effective method. 

Let us show the main idea of the approach on the example of the function
$F(x) = (2x+1)^{-10} x^{-3}$. 
Taking into account that  
$(1+\varepsilon)^\alpha=1+\alpha \varepsilon + \frac{1}{2!} \alpha(\alpha-1) \varepsilon^2 + \frac{1}{3!} 
\alpha(\alpha-1)(\alpha-2) \varepsilon^3 + \ldots $
we expand $F$ around $x=-1/2$ and $x=0$, 
choosing as local parameters respectively $\xi=2x+1$ and $\eta=x$. Then
\[
\frac{1}{\xi^{10}x^3} = \frac{1}{x^3}\ (1-20x+220x^2) + 
 \frac{8}{\xi^{10}}\ (1+3\xi+6\xi^2+10\xi^3+\ldots+
 45\xi^8+55\xi^9) + h(x) \ .
\] \par \noindent
The crucial point is that $h(x) \equiv 0$. Indeed, by construction $h$ is
bounded and regular both in $x=0$ and $x=-1/2$. Hence, by the
Liouville theorem we have $h(x)\equiv{\rm const}$. The constant is zero by the
 obvious boundary condition: $h \rightarrow 0$ for $|x| \rightarrow \infty$.
The presented approach simplifies 
computations in a striking way. 
We can integrate very quickly
any rational function with known poles. 

The Liouville theorem is the corner stone of the dressing method
(as pointed out by Its \cite{Its}): local behaviour of analytic functions
(e.g., around singularities) can define them globally.  However, the matrix
case is much more complicated (and more interesting) than the scalar one
(compare \cite{Ga,Sach}).

 In this paper we apply  methods of the theory of integrable systems to
the geometry of surfaces immersed in Euclidean spaces. The main idea of the 
{\bf soliton surfaces approach} consists in associating
with a given integrable system of nonlinear partial differential equations a
class of surfaces (or manifolds) immersed in the Lie algebra of the
corresponding linear problem using the so called Sym-Tafel formula
\refEq{Sym} (see \cite{Sym}, compare also \cite{Ci-FG,FG,Hu}).
In this way  we
can unify in a natural way various nonlinear models (spins, vortices, chiral
fields, etc.).

 We present several examples, mostly immersions in
${\bf R}^3$: surfaces swept out by various vortex motions, surfaces of constant
Gaussian curvature, constant mean curvature surfaces,
Bianchi surfaces, isothermic surfaces. 
The Sym-Tafel formula can be also applied to manifolds immersed in spaces of
higher dimensions like orthogonal coordinates in ${\bf R}^n$ \cite{Ci-cross}
and $n$-dim.\ space forms immersed in ${\bf R}^{2n-1}$ \cite{Ci-space}.

The {\bf Darboux-Bianchi} transformations (DBB transformations on the level of
surfaces) coincides, as a rule, with classical transformations studied by
Luigi Bianchi and other great geometers of XIX century  
\cite{Bob,Lamb,Sym}.

\section{The Darboux-B\"acklund transformation}
\label{D-BT}

There are several approaches to the construction of the Darboux matrix.
All these methods are closely related but usually not much attention
is given to this fact.
 The discussion of the relationship between the
Matveev method and the Zakharov-Shabat approach is a positive exception
(\cite{Bob-DB}, see also \cite{MS}). In this paper we compare  
the Zakharov-Shabat method with the approach of Neugebauer.

\subsection*{The Neugebauer-Meinel approach}

An especially effective method to construct $N$-fold Darboux-B\"acklund
transformation has been proposed by Neugebauer (\cite{MNS,NK,NM}, see
also \cite{Gu}).
The Darboux matrix is assumed to be a polynom in $\lambda$,
\begin{equation}  \label{D polynom}
     T = \sum_{j=0}^N C_j(x^1,x^2) \lambda^{j} 
\end{equation} \par \noindent		
(polynomial Darboux matrices will be denoted by $T$).
Any matrix $D$ rational in $\lambda$ can be replaced by an equivalent matrix
$T$ of the form \refEq{D polynom} (compare Remark~\ref{equivalent} where for $f$ 
we should take the least common denominator). One can consider
other equivalent matrices, e.g., polynoms in $1/\lambda$.

From the elementary linear algebra we know that $T^{-1} = (\det T)^{-1} T'$, 
where $T'$ is the matrix of cofactors of $T$.
Obviously $T'$ is a polynom in $\lambda$.
Therefore, if $U_k$ are rational functions of $\lambda$, then $\tilde{U_k}$ given by
\refEq{DBTU} are rational as well.  What is more, the only candidates for poles
of $\tilde{U}_k$ are poles of $U_k$ and zeros of $\det T$. 

\begin{lem} \label{det const}
If\ \ ${\rm Tr} U_k = 0$ then the determinant of the Darboux matrix does not
depend on $x^1, x^2$. In particular, the zeros of $\det T$ are constant.
\end{lem}

The proof is based on a very useful identity, $(\log \det A),_x = {\rm Tr} (A,_x
A^{-1})$, which holds for any nondegenerate matrix function $A=A(x)$. 
The condi\-tions ${\rm Tr} U_k = 0$, assumed throughout the paper, are not
particularly restrictive: practically all linear problems associated with
integrable systems  are traceless.

The necessary
condition for $T$ to be a Darboux matrix is the requirement that $\tilde{U}_k$ 
have no poles in zeros of $\det T$.
We assume that  $\det C_0 \neq 0$ and  all zeros of $\det T$ (denoted by
$\lambda_i$) are simple and pairwise different. The total number of zeros  is $Nn$.
We assume also that $U_1$, $U_2$ and $\Phi$  are regular in $\lambda_i$.
Because  $\tilde{U}_k=
\tilde{\Phi},_k \tilde{\Phi}^{-1}= \tilde{\Phi},_k \tilde{\Phi}' (\det\Phi \det
T)^{-1}$, then the
condition for $\tilde{U}_k$ to have no pole in $\lambda=\lambda_i$ is given by
\begin{equation} \label{residua}
     \tilde{\Phi},_k (\lambda_i) \tilde{\Phi}'(\lambda_i) = 0 \ .
\end{equation} \par \noindent
We proceed to construct $\tilde{\Phi}$ in order to satisfy the equations
\refEq{residua}.  The matrix $\tilde{\Phi}(\lambda_i)$ is degenerated. Hence there
exists a vector $p_i$ such that $\tilde{\Phi}(\lambda_i) p_i = 0$. 
Then, by \refEq{ZS} it follows that $\tilde{\Phi},_k (\lambda_i) p_i = 0$.
Differentiating the equation $\tilde{\Phi}(\lambda_i) p_i = 0$ we obtain
$\tilde{\Phi}(\lambda_i) p_i,_k = 0$.   Therefore, taking into
account that zeros $\lambda_i$ are simple, we obtain that
$p_i,_k$ is proportional to $p_i$. In fact, using an additional freedom ($p_i$
are defined up to a scalar factor) we can  choose $p_i$ to be constant
vectors.

\begin{cor}
If $\lambda_i$ is a simple zero of $\det T$, then there exists
a constant vector $p_i\neq 0$ such that 
$\tilde{\Phi}(\lambda_i) p_i = \tilde{\Phi},_k (\lambda_i) p_i = 0$.
\end{cor}

The corollary implies the equation \refEq{residua}. Indeed, one has
only take into account the following fact of the elementary linear algebra (we
suggest to prove it as an exercise) \cite{MNS}.

\begin{lem}
If there
exists a vector $p\neq 0$ such that 
$ X p = 0$  and  $ Y p = 0$ ($X,Y$ are matrices) then $YX'=0$.
\end{lem}

Finally, we can treat constants $\lambda_i\in {\bf C}$ and $p_i\in {\bf C}^n$
$(i=1,\ldots,nN)$ as prescribed parameters. Then the equations
$\tilde{\Phi}(\lambda_i) p_i = 0$ will determine the coefficients $C_j$ of the
Darboux matrix \refEq{D polynom}. Namely, the system
\begin{equation} 
     T(\lambda_k) \Phi(\lambda_k) p_k = 0 \ , \quad (k=1,\ldots, Nn) \ ,
\end{equation} \par \noindent
is equivalent to $Nn^2$ scalar equations for $(N+1)n^2$ scalar
coefficients. The remaining freedom corresponds  to the gauge freedom. For
instance we can treat $C_0$  as still
undetermined. Then matrices $C_1$,\ldots,$C_N$ are uniquely determined
provided that the complex parameters $\lambda_k$ and constant complex vectors $p_k$
are given. 

The {\bf degenerate case} (multiple zeros of $\det T$) is more complicated.
If there exist several vectors satisfying the equation $\tilde{\Phi}(\lambda_j) p_j =
0$, then they also can be chosen to be constant.

To obtain explicit solutions of a given nonlinear system
by the presented method we have to know explicitly at least
one solution of the linear problem \refEq{ZS}. Then we can obtain in
an algebraic way a sequence of other solutions which can be interpreted as
a ``nonlinear superposition'' of the given (``background'') solution and 
some number of solitons.
In particular, applying the transformation \refEq{D polynom} to the
trivial background we obtain ``pure'' $N$-soliton solutions.

\subsection*{The Zakharov-Shabat approach}

In the papers of Zakharov, Shabat and Mikhailov another representation of the
Darboux matrix was used \cite{Mi,ZMNP,ZM-ZETF,ZM-CMP},
\begin{equation}   \label{ZM}
D = {\cal N} \left( I + \sum_{k=1}^N \frac{A_k}{\lambda-\lambda_k} \right) \ , \qquad
D^{-1} = \left(I + \sum_{k=1}^N \frac{B_k}{\lambda-\mu_k} \right) {\cal N}^{-1}  \ ,
\end{equation} \par \noindent
where $I$ is the unit matrix, ${\cal N},A_k,B_k$ depend on $x^1,x^2$ and $\lambda_k,
\mu_k$ are constants (assumed to be pairwise different).

The similar form of $D$ and $D^{-1}$ is a restriction on $A_k$ and $B_k$.
The condition $D D^{-1}=I$ is equivalent to
\begin{equation}  \label{DD-1}
 A_k \left( I + \sum_{j=1}^N \frac{B_j}{\lambda_k-\mu_j} \right) = 0 \ , \qquad
 \left( I + \sum_{j=1}^N \frac{A_j}{\mu_k-\lambda_j} \right) B_k = 0 \ ,
\end{equation} \par \noindent
$(k=1,\ldots,N)$. In the case $N=1$ (``1-soliton case'') this system can be
easily solved to give
\begin{equation}  \label{DM generic}
     D = {\cal N}  \left(  I + \frac{\lambda_1-\mu_1}{\lambda-\lambda_1} P \right) \ , \qquad
     D^{-1} = \left(  I + \frac{\mu_1-\lambda_1}{\lambda-\mu_1} P \right) {\cal N}^{-1} \ , 
\end{equation} \par \noindent
where $P$ is a projector (i.e.\ $P^2=P$).

\begin{Th}[Zakharov, Shabat \cite{ZS}] If 
$P$ is given by the formulas:
\begin{equation}  \label{14}
     \ker  P = \Phi (\lambda_1) V_{ker} \ , \quad
     {\rm im}  P  = \Phi (\mu_1) V_{im} \ ,
\end{equation} \par \noindent
\noindent where   $V_{ker}$ and $V_{im}$ are
constant vector spaces such that $V_{ker} \oplus V_{im} = {\bf C}^n$,
then the transformation \refEq{DBTU} with $D$ given by \refEq{DM generic}
preserves the divisors of poles of matrices $U_k$.
\end{Th}

Every projector is completely characterized by its image and kernel.
The matrix 
of the projector $P$ is explicitly given by
\[
    P = \left\{ 0 , \Phi(\mu_1)V_{im} \right\} \left\{ \Phi(\lambda_1)V_{ker}, 
     \Phi (\mu_1) V_{im} \right\} ^{-1} \ , 
\] \par \noindent 
\noindent where  in the brackets we have $n\!\times\! n$ matrices. 
We use the same notation to designate a subspace of ${\bf R}^n$ and 
a matrix representing it (the columns of the matrix span the considered
vector space).

Multiplying $D$ given by \refEq{ZM} by the common denominator we obtain an
equivalent polynomial matrix like \refEq{D polynom}. However \refEq{ZM} and
\refEq{D polynom} are not strictly equivalent. One should remember about
restrictions \refEq{DD-1}. An equivalent matrix $D$ of the form \refEq{ZM}
exists only for some particular polynomial matrices $T$.

\section{Algebraic representation of the linear problem} \label{Reductions}
\label{ALG}

The techniques presented above guarantee that a linear problem
subject to the Darboux-B\"acklund transformation is transformed into
a linear problem of the same analytical dependence on $\lambda$.
However, a given integrable system  usually needs much more
restrictions imposed on the associated linear problem.
The most important reductions consist in confining $U_k$ to some Lie algebra.
Then $\Phi$ is confined to the corresponding group and, obviously, $D$ is
confined to the same group. An
example is in order.  Let $U_k^{\dag}(\overline{\lambda}) = - U_k(\lambda)$ for $k=1,2$
(this property holds in the case of the linear problem \refEq{NSE-problin} where
the coefficients by powers of $\lambda$ are ${\bf su}(2)$-valued). Then one can check
that $((\Phi^{\dag}(\overline{\lambda}))^{-1}),_k = U_k (\lambda)
(\Phi^{\dag}(\overline{\lambda}))^{-1}$ and, by virtue of Remark~\ref{C},
$(\Phi^{\dag}(\overline{\lambda}))^{-1}  = \Phi(\lambda) C(\lambda)$ where $C(\lambda)$ does not
depend on $x^1,x^2$. Elementary calculations show that $C^{\dag}(\overline{\lambda}) =
C(\lambda)$.  In fact $C$ depends only on the initial condition
$\Phi(x^1_0,x^2_0;\lambda)$ and by an appropriate choice of the condition we can
put $C=I$. Thus $\Phi$ satisfies $\Phi^{\dag}(\overline{\lambda}) \Phi(\lambda) = I$.
In general, if $U_k$ are  confined to some loop algebra 
then, by an appropriate choice of initial conditions, 
$\Phi$, $\tilde{\Phi}$ and $D$ can be restricted to the corresponding
group. 

The {\bf reduction groups} had already been introduced in the pioneering paper
of Zakharov and Shabat (\cite{ZS}, see also \cite{ZMNP}). Then the subject was
developed by Mikhailov \cite{Mi-group,Mi,ZM-ZETF,ZM-CMP}.
He classified a large class of reductions considering groups of 4 types
\cite{Mi-group}.

The restrictions on the matrix $D$ given by \refEq{DM generic} 
has been discussed in detail 
in \cite{Ci-dbt}. Actually, because of geometrical applications, it is
convenient to consider the unimodular Darboux matrix ${\cal D}$
\cite{Ci-dbt,Sym}  (by Lemma~\ref{det const} it is clear
that in the case of traceless linear problems one can always impose the
condition $\det D =1$).  To confine the Darboux matrix \refEq{DM generic} to the
group ${\bf SL}(n,{\bf C})$ it is sufficient to multiply it by the scalar coefficient
$f=(\det D)^{1/n}$:
\begin{equation}     \label{improved D}
     {\cal D} = \left( \frac{\lambda - \lambda_1}{\lambda - \mu_1} \right)^{d/n} {\cal N}  \left( I + 
          \frac{\lambda_1 - \mu_1}{\lambda - \lambda_1} P \right) \ ,                         
\end{equation} \par \noindent
where   $\det {\cal N} $=1,   $d$=dim(im$P$)   and   $\lambda_1  \neq  \mu_1$.
The matrices \refEq{DM generic} and  \refEq{improved D} are 
equivalent (compare Remark~\ref{equivalent}).
The paper \cite{Ci-dbt} contains the full list of restrictions
on on $\lambda_1, \mu_1, f, P, {\cal N} $ 
implied by the Mikhailov reductions provided that $\det {\cal N} \neq 0$.

Here we are going to present briefly some results concerning 
Darboux matrices in the polynomial form \refEq{D polynom}.

Let $U_k(-\lambda)=J U(\lambda) J^{-1}$ where $J$ can depend on $\lambda$. 
One can easily see that $J$ has to satisfy $J(-\lambda) J(\lambda) = \theta(\lambda) I$
where $\theta$ is a scalar function. The wave
function and the Darboux matrix can be confined to the group 
$\Phi(-\lambda) = J \Phi(\lambda) J^{-1}$, $T(-\lambda) = J T(\lambda) J^{-1}$.
Therefore $\det T(-\lambda) = \det T(\lambda)$ which means that zeros of $\det T(\lambda)$ 
appear in pairs $\lambda_j=-\lambda_k$. 
The corresponding constant eigenvectors $p_j$ and $p_k$ satisfy 
$\tilde{\Phi}(\lambda_k) p_k = 0$ and $\tilde{\Phi}(\lambda_j) p_j = 0$ which implies
$\tilde{\Phi}(\lambda_k) J(-\lambda_k) p_j = 0$. If the zeros $\lambda_j$ are simple
then $p_j = J(\lambda_k) p_k$.

The unitary reductions  
$U_k^{\dag} (\overline{\lambda}) = - H U_k (\lambda) H^{-1}$ are more
complicated. If we admit a (rational) dependence of $H$ on $\lambda$  then 
$H^{\dag}(\overline{\lambda}) = \eta(\lambda) H(\lambda)$ where $\eta$ is a scalar function.
One can easily check that $H ( \Phi^{\dag}(\overline{\lambda}) )^{-1}$
satisfies the system \refEq{ZS} provided that $\Phi$ is a solution.
$\tilde{\Phi}$ is subject to an analogical constraint. Thus we can derive
the condition 
\begin{equation} \label{DH}
 T(\lambda) = k(\lambda)  H \left( T^{\dag}(\overline{\lambda}) \right)^{-1} H^{-1} \ ,
\end{equation} \par \noindent
($k$ is a scalar function of $\lambda$) which is sufficient for $T$ to be the
Darboux matrix. 
We point out that now the simplest choice $k(\lambda) \equiv 1$ is 
not possible because it is incompatible with the form \refEq{D polynom}.
If $H$ is rational then $k$ has to be rational as well. From
\refEq{DH} we obtain
\begin{equation}  \label{k^n}
(k(\lambda))^n  = \overline{\det T(\overline{\lambda})} \det T(\lambda)  \ .
\end{equation} \par \noindent
Therefore $k(\lambda)$ is a polynom with real coefficients. We confine ourselves 
to the case of $k(\lambda)$ without real roots, i.e.
\[
k(\lambda) = (\lambda-\lambda_1)(\lambda-\overline{\lambda}_1)\ldots(\lambda-\lambda_N)(\lambda-\overline{\lambda}_N)
\] \par \noindent
where $N$ is the degree of the polynom $T(\lambda)$. Then  
\[
\det T =  (\lambda-\lambda_1)^{n-d_1}(\lambda-\overline{\lambda}_1)^{d_1}\ldots
          (\lambda-\lambda_N)^{n-d_N}(\lambda-\overline{\lambda}_N)^{d_N} \ .
\] \par \noindent
where $d_k$ are some integers. Note that for $n>2$ the zeros of $\det T$ 
are, as a rule, degenerated. If $\lambda_k$ are pairwise different, then dividing $T$ by
$(\lambda-\lambda_1)\ldots(\lambda-\lambda_N)$ we obtain the matrix $D$ (equivalent to $T$)
bounded for $\lambda \rightarrow \infty$:
\[
D = \frac{T(\lambda)}{(\lambda-\lambda_1)\ldots(\lambda-\lambda_N)} = S_0 + \sum_{k=1}^N 
\frac{S_k}{\lambda-\lambda_k} 
\] \par \noindent
where $S_k$ are some matrices dependent on $x^1,x^2$. From \refEq{k^n}
we can deduce that the matrix $S_0$ is not degenerated.
Therefore $D$ is exactly of the form \refEq{ZM}.
Using \refEq{DH} we compute $D^{-1}$ and, as a result we have 
$\mu_k = \overline{\lambda}_k$. The case of $k(\lambda)$ with real roots
is left as an exercise. We just mention that this case is possible
only for even $n$.

General conclusion is very interesting: in the important case of
unitary reductions the Darboux matrix can always be represented in
the form \refEq{ZM}.

\subsection*{Multilinear invariants of the Darboux matrix} \label{INVAR}

Let $\lambda_0$ be a pole  of matrices $U_1$ and $U_2$  
of the order $J$ and $K$ respectively.
The matrices $U:=(\lambda-\lambda_0)^J U_1$ and $W:=(\lambda-\lambda_0)^K U_2$ are holomorphic
in the neighbourhood of $\lambda_0$:
\[
  U = \sum_{i=0}^{\infty} u_i (\lambda-\lambda_0)^{i} \ , \quad
  W = \sum_{i=0}^{\infty} w_i (\lambda-\lambda_0)^{i} \ .
\] \par \noindent  
We assume that $u_j$ and $w_j$ are ${\bf sl}(n,{\bf C})$-valued functions of $x^1$,
$x^2$.  In many interesting cases including the equation \refEq{NSE}
the only singularity is
$\lambda_0=\infty$, i.e. $U_1$ and $U_2$ are polynomials in $\lambda$ while $U, W$ are
polynomials in $1/\lambda$.

The transformation \refEq{DBTU} for $U, W$ reads:
\begin{equation}  \label{DBT-UW}
	 \tilde{U} = (\lambda-\lambda_0)^J D,_1 D^{-1} + D U D^{-1} \ , \quad
    \tilde{W} = (\lambda-\lambda_0)^K D,_2 D^{-1} + D W D^{-1} \ .
\end{equation} \par \noindent
Considering the leading terms in equations \refEq{DBT-UW} we can obtain
a number of invariants of DBB transformation. In the sequel we assume
$J \leq K$ and $D$ is regular at $\lambda=\lambda_0$.

{\bf Linear invariants.}
Consider $\alpha U+ \beta W$ where $\alpha =\alpha(x^1,x^2)$ and $\beta=\beta(x^1,x^2)$
are given functions. We have
\[
     \alpha \tilde{U}+\beta \tilde{W} = \alpha (\lambda-\lambda_0)^J D,_1 D^{-1} + 
	  \beta (\lambda-\lambda_0)^K D,_2 D^{-1} + D(\alpha U+\beta W)D^{-1} \ .
\] \par \noindent
Let $L < J \leq K$. 
If $\alpha U + \beta W$ has zero of $L$-th order in $\lambda=\lambda_0$ 
then $\alpha \tilde{U}+\beta \tilde{W}$
has zero of at least the same order. Thus we have shown the following result
\cite{Ci-dbt}.

\begin{prop}  \label{linear invariants}
If $\lambda={\rm const}$ then any system of $L+1$ linear constraints ($L<J$)
of the form
\[
		\alpha u_m + \beta  w_m = 0 \ , \qquad (m=0,1,\ldots,L) \ ,
\] \par \noindent
where $\alpha =\alpha (x^1,x^2)$, $\beta=\beta(x^1,x^2)$ are given functions, 
is preserved by DBB transformation. 
\end{prop} 

The proposition can be extended for $L=J$ provided 
that the Darboux matrix satisfies the condition $D,_1(\lambda_0)=0$
(in the case $\lambda_0=\infty$ it means that the
normalization matrix does not depend on $x^1$).

In the case of the nonlinear Schr\"odinger equation we have (by inspection):
$u_0 + 2 w_0=0$ and $u_1 + 2 w_1=0$ (see \refEq{NSE-problin}). These constraints
are invariants of DBB transformation provided that ${\cal N} ={\cal N} (x^2)$.

{\bf Bilinear invariants.}
Let the center dot denotes an invariant scalar product in ${\bf sl}(n,{\bf C})$,
namely  $a\cdot b := {\rm Tr}(ab)$. 
Consider the transformation of $W\cdot W$:
\[
   \tilde{W} \cdot \tilde{W} = W \cdot W + 2 (\lambda-\lambda_0)^K (D^{-1} D,_2) \cdot W +
	(\lambda-\lambda_0)^{2K} (D^{-1} D,_2) \cdot (D^{-1} D,_2) \ .
\] \par \noindent
On the right hand side the leading terms (of order less than $K$) are contained
in $W\cdot W$.

\begin{prop} \label{bilinear invariants}
If $\lambda={\rm const}$  then the constraint 
\[
   \sum_{k=0}^m w_k \cdot w_{m-k} = \gamma_m  \qquad (0<m<K-1) 
\] \par \noindent
(where $\gamma_m = \gamma_m(x^1,x^2)$ is a given function) is preserved
by DBB transformation.
\end{prop}

The proposition holds for $m=K$ if and only if $(D^{-1}(\lambda_0) D,_2 (\lambda_0))
\cdot w_0 =0$  (in the polynomial case ($\lambda_0=\infty$) this condition 
reduces to ${\cal N},_2=0$).

The multilinear invariants can be used to obtain reductions of the linear
problem \refEq{ZS}. The reductions are fixed by the choice of 
$\alpha,\beta$ and $\gamma_m$.

 Similar results can be obtained considering  
the expansions of $U\cdot U$ and $U\cdot W$ around  $\lambda=\lambda_0$. One can
expect to get analogical results for ${\rm Tr}(U U \ldots U)$,
${\rm Tr}(W W \ldots W)$, ${\rm Tr}(U W \ldots W)$ etc.

\subsection*{The linear problem as a system of algebraic constraints on two
matrices}

Let us summarize the results of the preceding sections.
The knowledge of the linear problem like \refEq{ZS} is crucial to solve
a given integrable system.
One can describe the linear  problem parameterizing it explicitly by
soliton fields, their derivatives and integrals. However, if the
the construction of the Darboux matrix is concerned, much more convenient 
description is by a system of relatively simple (non-differential)
constraints ({\bf algebraic representation of the linear
problem} \cite{Ci-dbt,Ci-ChSF}): matrices $U_1$ and $U_2$  are rational functions of $\lambda$ (with
prescribed poles) restricted to some loop algebra (in other words, their
coefficients by powers of $\lambda$ are restricted to a Lie algebra) and some
multilinear constraints are imposed on the coefficients by powers of $\lambda$.
Considering nonisospectral linear problems \cite{BZM} we should also prescribe
an explicit dependence of variable spectral parameter on a constant parameter
\cite{Ci-dbt}.

     The algebraic representation of the linear  problem  allows  us  to
introduce a very useful {\bf definition of the Darboux matrix}. 
 The  matrix  $D$   is  called  the  Darboux   matrix for the linear problem
\refEq{ZS} if the transformation \refEq{DBTU} preserves the system of algebraic 
constraints equivalent to this linear problem \cite{Ci-dbt}.

	It is very convenient to divide the construction of the Darboux  matrix  
into two separate steps (\cite{Ci-dbt,Ci-ChSF}, similar idea can
be found also in \cite{Its}).
 First, we represent the linear
problem under consideration as a system of algebraic 
constraints on two matrices. 
This step may turn out to be rather non-standard (i.e.\ each case has to be
treated individually) but involves no cumbersome calculations. Usually the
algebraic constraints can be found by straightforward inspection of the given
linear problem. The crucial point is to find the reduction group.
Second, we derive the constraints on the Darboux matrix imposed by 
the requirement  that the matrix has to preserve the algebraic constraints.
The second step is technically much more difficult but there exist many 
general propositions which
make the construction practically algorithmic (the 1-soliton case is
discussed in detail in \cite{Ci-dbt}).

\section{Soliton surfaces approach}
\label{SYM}

We proceed to apply the soliton theory to the geometry of surfaces immersed
in Euclidean spaces.
Let us consider a surface $\Sigma \subset {\bf R}^3$ equipped (locally) with
coordinates $x^1$, $x^2$, and defined by the position vector
${\bf r}={\bf r}(x^1,x^2)\in{\bf R}^3$. 
The first fundamental form (``metric tensor'') is given by $I:= d{\bf r}\cdot 
d{\bf r}$
(the center dot means a scalar product in the ambient space ${\bf R}^3$).
Therefore, in local coordinates, $I:=g_{ij} dx^i dx^j$ (we assume the Einstein
convention, i.e., summation over repeating indices), where
$ g_{ij} := {\bf r},_i \cdot {\bf r},_j$.
The matrix inverse to $(g_{ij})$ will be denoted by $(g^{ij})$ and
$g:=\det (g_{ij})$.  Let ${\bf n}$ be
the unit normal vector,
${\bf n} := g^{-1/2} {\bf r},_1 \!\times\!\ {\bf r},_2 $,
where the cross denotes skew (or vector) product in ${\bf R}^3$.
The second fundamental form is defined as $II:=-d{\bf n} \cdot d{\bf r}$, or
equivalently, $II=b_{ij} dx^i dx^j$ where $ b_{ij} := {\bf r},_{ij} \cdot {\bf n}$.
The matrix with coefficients $b^i_j := g^{ik} b_{kj}$  represents the
``shape operator'' \cite{Boo}. The principal curvatures, $k_1$ and $k_2$, are 
defined as eigenvalues of $(b^i_j)$. The corresponding eigenvectors
(``principal directions'') can also be characterized by the requirement to
diagonalize both fundamental forms. 
The Gaussian curvature is defined as $K := k_1 k_2$ and the mean curvature
as $H:=\frac{1}{2}(k_1+k_2)$. They can be expressed by the fundamental forms:
$K = {\det(b_{ij})}/{\det(g_{ij})}$, $H= g^{ij} b_{ij}$.
In fact, $K$ can be expressed solely in terms of $g_{ij}$ (the Gauss ``Theorema
Egregium'').

Differentiating tangent vectors we obtain {\bf Gauss-Weingarten (GW)} equations
${\bf r},_{ij} = \Gamma^{k}_{ij} {\bf r},_k + b_{ij} {\bf n}\ $, where $\Gamma^{k}_{ij}$ are
some $x^1,x^2$-dependent coefficients known as ``Christoffel symbols''. It is
well known that they can be expressed entirely by $g_{ij}$, namely
$\Gamma^{k}_{ij} = \frac{1}{2} g^{ks} ( g_{is},_j + g_{sj},_i - g_{ij},_s)$.
The GW equations can be rewritten as kinematic equations of the
orthogonal frame ${\bf e}_1$, ${\bf e}_2$, ${\bf n}$, where ${\bf e}_k$ form an orthogonal basis
in the tangent space:
\begin{equation}  \label{GW-matrix}
\frac{\partial}{\partial x^k} \left( \begin{array}{c} {\bf e}_1 \\ {\bf e}_2 \\ {\bf n} \end{array} \right) = 
\Omega_k \left( \begin{array}{c} {\bf e}_1 \\ {\bf e}_2 \\ {\bf n} \end{array} \right) \ , 
\end{equation} \par \noindent
where $\Omega_1$ and $\Omega_2$ are {\bf so}(3) matrices depending on $g_{ij}$ and
 $b_{ij}$.
It is convenient to use the well known isomorphism ${\bf so}(3) \simeq {\bf su}(2)$
(see, for instance, \cite{Cor}) and to rewrite the equations \refEq{GW-matrix}
in terms of $2\!\times\! 2$ matrices. The explicit form of the matrices reads,
for example, as follows (compare \cite{Ci-FG,S-6}):
\begin{equation} \label{GW-su2}
	\Phi,_k=\frac{i}{2\sqrt{g_{11}}} \left( \begin{array}{cc} \frac{g_{11} b_{2k} - 
g_{12} b_{1k}}{\sqrt{g}} & b_{1k} + i \Gamma^2_{1k} \sqrt{\frac{g}{g_{11}}} \\
b_{1k} - i \Gamma^2_{1k} \sqrt{\frac{g}{g_{11}}} & \frac{g_{12} b_{1k} - 
g_{11} b_{2k}}{\sqrt{g}} \end{array} \right) \Phi
	\qquad (k=1,2)\ .
\end{equation} \par \noindent
Integrability conditions for GW equations 
can be derived from ${\bf r},_{ijk}={\bf r},_{ikj}$. They are expressed explicitly in
terms of matrices $\Omega_k$:
\[
   \Omega_1,_2 - \Omega_2,_1 + [\Omega_1, \Omega_2] = 0 \ ,
\] \par \noindent
and are equivalent to a system of 3 nonlinear equations 
 for the coefficients
$g_{ij}$, $b_{ij}$: the 
{\bf Gauss-Mainardi-Co\-daz\-zi (GMC)} equations.

\subsection*{The Sym-Tafel formula}  \label{SolSurf}

Let  matrices $U_1$, $U_2$ for $\lambda \in {\bf R}$ assume values in a Lie algebra $g$
(in other words, matrices $U_k$ are meromorphic functions of $\lambda$ with
$g$-valued coefficients). Then one can confine $\Phi$ to the corresponding Lie
group which implies that the matrix ${\bf r}={\bf r}(x^1,x^2;\lambda)$ given by 
\begin{equation}    \label{Sym} 
     {\bf r} = \Phi ^{-1}\Phi ,_\lambda 
\end{equation} \par \noindent
is $g$-valued. The formula \refEq{Sym} is known as the Sym formula or 
the Sym-Tafel formula (the final step in derivation of this formula was 
done by Tafel, see \cite{S-1}).  We assume that
the Lie algebra $g$ is equipped with a scalar product invariant with respect to
adjoint transformations, $\scalarproduct{Ad_\Phi a}{Ad_\Phi b} =
\scalarproduct{a}{b}$ (in a matrix representation $Ad_\Phi a := \Phi^{-1} a
\Phi$). A well known example is the Killing-Cartan form, $\scalarproduct{a}{b}
:= {\rm Tr} (ad_a \circ ad_b)$, which is non-degenerate for semi-simple
algebras. In general,  ${\rm Tr}(f(a) f(b))$ defines an invariant bilinear form
for any matrix representation $f$ of the Lie algebra $g$, and can be used as a
scalar product provided that the bilinear form is non-degenerate \cite{GG}.
Then $g$ can be identified with a pseudo-Euclidean space ${\bf R}^m$ and the
function ${\bf r}$ represents a $\lambda$-family of parametric surfaces 
\cite{Th} ({\bf ``soliton surfaces''}) immersed in $g \cong  {\bf R}^m$.
Kinematics of ${\bf r}$ defines an integrable nonlinear model which can be
interesting in itself
\cite{S-2,Sym}.

Starting from the formula \refEq{Sym} we compute tangent vectors
${\bf r},_k =  \Phi^{-1} U_k,_\lambda \Phi$  and the metric tensor 
$g_{ij} = \scalarproduct{{\bf r},_i}{{\bf r},_j} = \scalarproduct{U_i,_\lambda}{U_j,_\lambda}$  (the
invariance of the scalar product was used).
We obtained $g_{ij}$ without solving differential
equations \refEq{ZS}. It was sufficient to know only the matrices $U_1, U_2$.
GW equations read 
${\bf r},_{kj} =  \Phi^{-1} (U_k,_{\lambda j} + [U_k,_\lambda,\ U_j] ) \Phi$ and
GMC equations (${\bf r},_{ijk}={\bf r},_{ikj}$) are given by 
$ \Phi^{-1} [U_i,_{\lambda},\ U_j,_k - U_k,_j + [U_j,\ U_k] ] \Phi = 0$.

The Sym-Tafel formula gives an interesting connection between the classical
geometry of manifolds (with possible singularities) immersed in ${\bf R}^m$ and the
theory of solitons \cite{Sym}. 
The soliton surfaces approach is very useful in construction of the
so called ``integrable geometries'' \cite{Bob,Ci-Gall,CGS-izot,Sym}.
Indeed, any class of soliton surfaces given by \refEq{Sym} is 
integrable. 
 Geometrical
objects  associated  with  soliton surfaces (tangent vectors, normal vectors,
foliations  by  curves etc.) usually can be identified with solutions to  some
nonlinear models (spins, chiral models, strings, vortices etc.) \cite{S-2,Sym}.
The transformation \refEq{DBTPhi} can be immediately extended on the models
associated with soliton surfaces.  For  instance,  the  transformation
\refEq{DBTPhi}  applied  to   the position vector \refEq{Sym} assumes the form
\begin{equation}      \label{r}
     \tilde{\bf r} = {\bf r} +  \Phi ^{-1} {\cal D}^{-1} {\cal D},_\lambda \Phi  
\end{equation} \par \noindent
and we propose to call it the {\bf Darboux-Bianchi transformation}.
Usually ${\bf r}\in g \subset {\bf sl}(n,{\bf C})$  and we may consider  
the unimodular Darboux matrix \refEq{improved D}. In the ${\bf SU}(n)$ case
we have the additional restriction $\mu_1=\overline{\lambda}_1$.
The transformation \refEq{r} for soliton surfaces 
immersed in ${\bf su}(n)$ is given by \cite{S-2}:
\begin{equation}     \label{-11}
   \tilde{\bf r} = {\bf r} + \frac{2 {\rm Im} \lambda_1}{
	\mbox{$\mid \lambda-\lambda_1 \mid$}^2} 
	\Phi ^{-1} i\left( \frac{d}{n}  - P \right) \Phi  \ ,
\end{equation} \par \noindent
\noindent where  $\lambda_1$  is a complex parameter and $P$ is the  hermitean  projector 
(i.e., $P^2=P$ and $P^{\dag}=P$) onto the space $\Phi (\overline{\lambda }_1 )V_{im}$,
where $V_{im}$ is a constant vector space of a given dimension $d$.
The length of the segment  $\tilde{\bf r}-{\bf r}$ does not depend on $x^1,x^2$.

\subsection*{Geometry from spectral problems}

The Euclidean 3-dimensional space ${\bf E}^3$ 
can be identified with the Lie algebra ${\bf su}(2)$ endowed with
the scalar product given by the Killing-Cartan form
\begin{equation}   \label{Trab}
    a \cdot b := \scalarproduct{a}{b} =  - 2 {\rm Tr} (a b) \ .
\end{equation} \par \noindent
Note that ${\rm Tr}(a^2) \leq 0$ for $a\in {\bf su}(2)$.
What is more, the skew product in ${\bf R}^3$ can be identified with the 
commutator of ${\bf su}(2)$ matrices.
The standard basis, orthonormal with respect to \refEq{Trab}, is given by
\begin{equation}  \label{su(2) basis}
     {\bf e}_1 = - \frac{1}{2} i \left( \begin{array}{cc} 0 & 1 \\ 1 & 0 \end{array} \right)  , \quad  
     {\bf e}_2 = - \frac{1}{2} i \left( \begin{array}{cc} 0 & -i \\ i & 0 \end{array} \right)  , \quad    
     {\bf e}_3 = - \frac{1}{2} i  \left( \begin{array}{cc} 1 & 0 \\ 0 & -1 \end{array} \right)  .
\end{equation} \par \noindent
In other words, we identify 
\[
      {\bf su}(2) \ni \frac{1}{2} \left( \begin{array}{cc} -iz & -y-ix \\ y-ix & iz \end{array} \right) 
\quad		\longleftrightarrow \quad  (x,\ y,\ z) \in {\bf E}^3 \ .
\] \par \noindent

Below we present several examples of integrable geometries in ${\bf E}^3$.  We start
from given linear problems and use the formula \refEq{Sym}.  The normal vector
to ${\bf r}$ is given by  ${\bf n} = (-2{\rm Tr}([U_1,_\lambda, U_2,_\lambda]^2))^{-1/2} \Phi^{-1}
[U_1,_\lambda\ U_2,_\lambda]\Phi$.  Then 
  we compute the fundamental forms and the mean or Gaussian curvature
of the resulting surface.

In order to put the result into a more elegant form we have chosen a special 
dependence between $\lambda$ and the spectral parameter $\zeta=\zeta(\lambda)$.
In two cases (constant mean curvature surfaces and spherical surfaces)
the linear problem depends on the spectral parameter through trigonometric
functions $\sin\! \kappa$ and $\cos\!\kappa$, where $\kappa$ is proportional to $\lambda$. 
It is equivalent to using the spectral parameter $\zeta$ confined to
the unit circle, i.e., $\sin\!\kappa = \frac{1}{2i} (\zeta-\frac{1}{\zeta})$ and
$\cos\!\kappa = \frac{1}{2} (\zeta+\frac{1}{\zeta})$.
We use notation  $x^1\equiv u$, $x^2\equiv v$.

\begin{itemize}
\item {\bf Pseudospherical surfaces}

\begin{equation} \begin{array}{c} \label{problin-pseudospherical}
		\Phi,_1 = (- \zeta {\bf e}_3 - \varphi,_1 {\bf e}_2) \Phi \ , \\[1ex]
		\Phi,_2 = {\zeta}^{-1} (\cos\!\varphi\ {\bf e}_3 - \sin\!\varphi\ {\bf e}_1) \Phi \ ,
\end{array}  \end{equation} \par \noindent
where $\zeta=\exp(-R\lambda)$. The compatibility conditions \refEq{cc} are equivalent to
the sine-Gordon equation $\varphi,_{12} = \sin\!\varphi$ and the fundamental forms are
given by
\begin{equation}   \begin{array}{c}  \label{ff-pseudospherical}
    I = R^2 ( \zeta^2 du^2 + 2 \cos\!\varphi\, du dv + \zeta^{-2} dv^2) \ , \\[1ex]
    II = 2 R \sin\!\varphi\, du dv \ .
\end{array}  \end{equation} \par \noindent
The Gaussian curvature is given by $K=-R^2$ (compare \cite{S-2,Sym}).

\item {\bf Constant mean curvature surfaces}

\begin{equation}  \label{problin-Hconst}
\begin{array}{c}  
    \Phi,_1 = \left( - (e^{\varphi/2} + e^{-\varphi/2} \cos\! 2\kappa) {\bf e}_1 
	 - \frac{1}{2} \varphi,_2 {\bf e}_2 - e^{-\varphi/2} \sin\! 2\kappa {\bf e}_3 \right) \Phi \ , \\[1ex]
	 \Phi,_2 = \left( - e^{-\varphi/2} \sin\! 2\kappa\  {\bf e}_1 + \frac{1}{2} \varphi,_1 {\bf e}_2 
	  - (e^{\varphi/2} - e^{-\varphi/2} \cos\! 2\kappa) {\bf e}_3 \right) \Phi \ ,
\end{array} \end{equation} \par \noindent
where $\kappa:=\frac{1}{2} H^{-1} \lambda$. GMC eqs. are reduced to the single equation 
\begin{equation}  \label{SHG}
        \varphi,_{11}+\varphi,_{22}+4\sinh\!\varphi = 0 \ ,
\end{equation} \par \noindent		  
known as the  elliptic sinh-Gordon equation. The fundamental forms of \refEq{Sym} read
\begin{equation} \begin{array}{c}  \label{ff-Hconst}
  I = H^{-2} e^{\varphi} (du^2 + dv^2) \ , \\[1ex]
  II = H^{-1} \left( (e^{\varphi}+\cos\! 2\kappa) du^2 + 2\sin\! 2\kappa \ du dv +
     (e^{\varphi}-\cos\! 2\kappa) dv^2  \right) \ ,
\end{array} \end{equation} \par \noindent
which  means that the surface has the constant mean curvature $H$ \cite{DS}.

\item {\bf Spherical surfaces}

\begin{equation} \begin{array}{c}    \label{problin-spherical}
    \Phi,_1 = \left( - 2\sinh\!\frac{\varphi}{2} \cos\!\kappa\ {\bf e}_1 
  - \frac{1}{2} \varphi,_2 {\bf e}_2 + 2\cosh\!\frac{\varphi}{2} \sin\!\kappa\ {\bf e}_3 \right) \Phi\ ,\\[1ex]
    \Phi,_2 = \left( - 2\sinh\!\frac{\varphi}{2} \sin\!\kappa\ {\bf e}_1 
  + \frac{1}{2} \varphi,_1 {\bf e}_2 - 2\cosh\!\frac{\varphi}{2} \cos\!\kappa\ {\bf e}_3 \right) \Phi \ ,
\end{array} \end{equation} \par \noindent
where $\kappa:=R\lambda$, and GMC eqs. are equivalent to \refEq{SHG}. 
\begin{equation} \begin{array}{c}    \label{ff-spherical}
    I = 2R^2 \left( (\cosh\!\varphi + \cos\! 2\kappa)\ du^2 + 2 \sin\! 2\kappa\ du dv 
	     + (\cosh\!\varphi-\cos\! 2\kappa)\ dv^2 \right) \ ,        \\[1ex]
	 II = 2R \sinh\!\varphi\ (du^2 + dv^2) \ .
\end{array} \end{equation} \par \noindent
Therefore, ${\bf r}$ defines a spherical surface ($K=R^{-2}>0$) \cite{DS}.

\item {\bf Bianchi surfaces}  \label{BIANCHI}

\begin{equation} \begin{array}{c} \label{problin-Bianchi}  
	\Phi ,_1 = \left(- \zeta a {\bf e}_3 - \left(\varphi,_1 + 
	\frac{\rho,_2 a}{2\rho b} \sin\!\varphi \right) {\bf e}_2 \right) \Phi \ , \\[2ex]
     \Phi ,_2 = \left( \frac{1}{\zeta} (b\cos\!\varphi\ {\bf e}_3 - b\sin\!\varphi\ {\bf e}_1) +
	  \frac{\rho,_1 b}{2\rho a} \sin\!\varphi\,{\bf e}_2 \right) \Phi \ ,
\end{array} \end{equation} \par \noindent	  
\noindent where  $\rho = f(u) + g(v)$ ($f,g$ are given functions) and
\[
     \zeta  = \left( \frac{1-2\lambda g(v)}{1+2\lambda f(u)} \right)^{1/2} \ , 
	  \qquad (\lambda ={\rm const}) \ ,
\] \par \noindent
is ``variable spectral parameter''. We obtain
\begin{equation}  \begin{array}{c}  \label{ff-Bianchi}
     I = (\zeta,_{\lambda})^2 \left( a^2\,du^2 + 2ab \zeta^{-2}\cos\!\varphi\,   
du dv + b^2 \zeta^{-4} dv^2  \right) \ , \\[1ex]
     II = - 2 \zeta,_{\lambda} \zeta^{-1} ab \sin\!\varphi\, du dv \ .
\end{array}  \end{equation} \par \noindent
The Gaussian curvature is $K = - \rho^{-2}$ where $
   \rho := \frac{f(u)}{1+2\lambda f(u)} + \frac{g(v)}{1-2\lambda g(v)}$. For $\lambda=0$
we have $\rho=f+g$ and the formulas \refEq{ff-Bianchi} assume the standard form:
\begin{equation}  \begin{array}{c}  \label{ff-Bianchi-standard}
I = \rho^2 \left( a^2\,du^2 + 2ab \cos\!\varphi\,du dv + b^2\,dv^2 \right)\ , \\[1ex]
 II = 2\rho ab \sin\!\varphi\,du dv \ .
\end{array}  \end{equation} \par \noindent	
We recognize the fundamental forms of Bianchi surfaces in asymptotic
coordinates (for more details see \cite{Ci-dbt,Kor,LS,Ta}).
\end{itemize}

\subsection*{Localized Induction Equations and 
multi-soliton surfaces in ${\bf R}^3$}

Soliton surfaces in ${\bf R}^3$ and several nonlinear models of physical 
importance \cite{CGrS,LSW-vortex,S-5,SRLB,Sym,CSW} are
associated with ${\bf su}(2)$ algebra. The corresponding linear problems
(known as ${\bf su}(2)$-AKNS linear problems \cite{AS}, see also \cite{Sym})  
are parameterized by analytic functions $\omega=\omega(\lambda)$.
A typical example is given  by the linear problem \refEq{NSE-problin} of the
nonlinear Schr\"odinger equation ($\omega(\lambda)=-2\lambda^2$).
The function $\omega$ uniquely characterizes the asymptotic behaviour of 
multi-soliton solutions. The linear problem 
corresponding to the trivial solution ($q\equiv 0$) has the form
\[
\Phi_0,_x = i\lambda \sigma_3 \Phi_0 \ ,\qquad \Phi_0,_t = i\omega(\lambda) \sigma_3 \Phi_0 \ ,
\] \par \noindent 
and can be solved easily
\begin{equation} \label{triv}
\Phi_0 = \exp (i\lambda x \sigma_3 + i \omega(\lambda) t \sigma_3) \ .
\end{equation} \par \noindent
The case $\omega=-2\alpha\lambda^2-4\beta\lambda^3$
is associated with the Hirota equation
\begin{equation} \label{NLS} 
iq,_2 + \alpha (q,_{11} + 2\mbox{$\mid q \mid$}^2 q) - i\beta (q,_{111} + 
6\mbox{$\mid q \mid$}^2 q,_1) = 0 \ , 
\end{equation} \par \noindent
where $\alpha,\beta$ are real constants and $q=q(x^1,x^2)\in {\bf C}$.
The kinematics of the position  vector to soliton surfaces (evaluated
at $\lambda=0$) of the Hirota
equation describes the motion of the single thin vortex
filament in the so called {\bf localized induction approximation with axial
flow} \cite{FM,KI}:
\begin{equation} \label{axial} 
 		{\bf r},_2= \alpha ({\bf r},_1\!\times\!\,{\bf r},_{11}) + \beta ({\bf r},_{111} + 
		\frac{3}{2} ({\bf r},_{11})^2 {\bf r},_1) \ ,  \qquad  ({\bf r},_1)^2= 1 \ , 
\end{equation} \par \noindent
where ${\bf r}={\bf r}(x^1 ,x^2) \in {\bf R}^3$. 
 The tangent vector $S:={\bf r},_1$ solves the following
spin model
\[
 S,_2 = \alpha (S \!\times\! S,_{11}) + \beta \left( S,_{11} + \frac{3}{2} (S,_1^2 S),_1 
\right) \ .
\] \par \noindent
The special case $\beta=0$ corresponds to Localized Induction Equation,
nonlinear Schr\"odinger equation and continuum Heisenberg ferromagnt model 
respectively (see, for instance, \cite{FT,LP,Ri,S-2}).
In the case $\alpha=0$, $q\in {\bf R}$ the equation \refEq{NLS} is known as modified
Korteweg-de Vries equation. Soliton surfaces are degenerated: the position
vector ${\bf r}$ describes the evolution of a plane curve  which  has  an
interesting elastomechanical interpretation and admits so called ``loop
solitons'' as special solutions \cite{S-5}.

 In the ${\bf su}(2)$ case usually 
the formula \refEq{Sym} is used with with the factor $\frac{1}{2}$. Therefore, in this
section we assume ${\bf r}=\frac{1}{2} \Phi^{-1}\Phi,_\lambda$.
Iterating the Darboux-B\"acklund transformation $N$ times we obtain 
the following expression for $N$-soliton surfaces (more precisely, $N$-soliton
addition to any background surface) associated with ${\bf su}(2)$-linear problems
\cite{Ci-Nsol,Sym}:
\begin{equation}  \label{iter}
   {\bf r}_{B+N} = {\bf r}_B + \sum_{k=1}^N d_k 
  \left( \frac{2{\rm Re} \Xi_k}{\mbox{$\mid \Xi_k \mid$}^2+1} {\bf e}_1 - 
  \frac{2{\rm Im} \Xi_k}{\mbox{$\mid \Xi_k  \mid$}^2+1}
 {\bf e}_2 + \frac{\mbox{$\mid \Xi_k  \mid$}^2-1}{\mbox{$\mid \Xi_k  \mid$}^2 
 +1} {\bf e}_3 \right)  \  , 
\end{equation} \par \noindent	
where the subscript  $B$ means ``background'', ${\bf e}_k$ are defined
by \refEq{su(2) basis}, $d_k$ are given by
\begin{equation} \label{dk}
    d_k := \frac{{\rm Im} \lambda_k}{\mbox{$\mid \lambda-\lambda_k \mid$}^2} \ ,
\end{equation} \par \noindent
$\lambda_k$ are constant complex parameters, and $\Xi_k$ parameterize $P_k$, 
orthogonal projectors onto 1-dim.\ subspaces of ${\bf C}^2$, defined
by ${\rm im} P_{k+1} = \Phi_{B+k}^{-1}(\lambda) \Phi_{B+k}(\overline{\lambda}_k) p_k$
where $p_k \in {\bf C}^2$ are constant vectors. Namely,
\[
	P_k = \frac{1}{1+|\Xi_k|^2} \left( \begin{array}{cc} |\Xi|^2 -1 & \Xi \\ \overline{\Xi} & 1-|\Xi|^2
  \end{array} \right) \ .
\] \par \noindent
Trying to compute ${\bf r}_{B+N}$ one meets serious technical problems even for
$N=2$ (compare \cite{LSW-vortex}). However, it is possible to simplify the
problem and to express 
 the formula \refEq{iter} in terms of 
functions $\xi_k$ ($k=1,2,\ldots,N$) defined by 
\begin{equation}  \label{xi}
 \xi_k := \frac{u_{k1}}{u_{k2}} \ , \quad  {\rm where} \quad
\left( \begin{array}{c}  u_{k1} \\ u_{k2} \end{array} \right) := \Phi_B^{-1}(\lambda) \Phi_B (\overline{\lambda}_k) 
 \left( \begin{array}{c} p_{k1} \\ p_{k2} \end{array} \right) \ ,
\end{equation} \par \noindent
where $p_{k1}, p_{k2}$ are components of $p_k$ (see \cite{Ci-Nsol}).
Let us represent the complex function $\xi_k$ by 
\begin{equation}  \label{exp}
      \xi_k =: \exp(Q_k - i\alpha_k) \ . 
\end{equation} \par \noindent		
where $Q_k$ and $\alpha_k$ are real functions. 
Then we can rewrite the formula \refEq{iter} for $N=1$ in a more
explicit way:
\begin{equation}  \label{N=1}
		{\bf r}_{B+1} = {\bf r}_B + \frac{d_1}{\cosh\! Q_1} \left( \begin{array}{c} \cos\! \alpha_1 \\  \sin\!\alpha_1 \\
		\sinh\! Q_1 \end{array} \right)  \ .
\end{equation} \par \noindent	
The representation \refEq{exp} is
especially convenient in the ${\bf su}(2)$-AKNS case. If $\Phi_B \equiv \Phi_0$
(see \refEq{triv}) then $Q_k$ and $\alpha_k$ are linear in $x$ and $t$
($x \equiv x^1$, $t \equiv x^2$):
\begin{equation} \label{Qal} \begin{array}{c}
   Q_k = 2 x {\rm Im} \lambda_k + 2 t {\rm Im} \omega_k + Q_{k0} \ , \\[1ex]
	\alpha_k = 2 x (\lambda - {\rm Re} \lambda_k) + 2 t (\omega - {\rm Re} \omega_k) + \alpha_{k0} \ ,
\end{array} \end{equation} \par \noindent
where $\omega:=\omega(\lambda)$, $\omega_k:=\omega(\lambda_k)$ and $Q_{ko}$, $\alpha_{k0}$
are constant. 
The soliton surface ${\bf r}_0$ degenerates to the straight line, 
${\bf r}_0 = -(x+\omega'(\lambda)t) {\bf e}_3$.
Therefore the formula \refEq{N=1}, valid for any background, is especially useful
in the case of the trivial background \refEq{triv} (${\bf r}_B \equiv {\bf r}_0$, 
$Q_1$ and $\alpha_1$ are given by \refEq{Qal}).
In this case ${\bf r}_{B+N}$, 
denoted by ${\bf r}_{N}$, describes the interaction of $N$ solitons.
A single soliton in the LIA case, $\omega(\lambda)=-2\lambda^2$,
has been first found by Hasimoto \cite{Ha}.

{\bf Physical characteristics} of the single soliton solution ${\bf r}_{1}$
can be computed in the standard way. First of all we determine
the maximum of the wave envelope ($Q_1=0$) which performs a helical 
movement. The {\bf group velocity} $v^g_1$ of
the soliton is computed as the velocity  of the maximum
along ${\bf e}_3$ axis. The rotation rate of the maximum is denoted by 
$\Omega_1$. Then we determine positions of the individual wave peaks
($\alpha_1=0$). Their velocity (or the {\bf phase velocity}) is almost constant
sufficiently far from the envelope maximum.  The phase velocity of the single
soliton wave will be denoted by $v^{ph}_1$. Straightforward computations yield:
\begin{equation} \label{phys} \begin{array}{c}
  v^g_k = \frac{{\rm Im} \omega_k}{{\rm Im} \lambda_k} - \omega'(\lambda) \ , \qquad
v^{ph}_k = \frac{\omega(\lambda) - {\rm Re} \omega_k}{\lambda-{\rm Re} \lambda_k} - \omega'(\lambda) 
\ , \\[2ex]
 \Omega_k = 2(\omega(\lambda) - {\rm Re} \omega_k) - 2(\lambda-{\rm Re} \lambda_k) 
 \frac{{\rm Im} \omega_k}{{\rm Im} \lambda_k} \ .
\end{array} \end{equation} \par \noindent
To describe interactions of solitons (the case $N>1$)
 it is convenient to introduce parameters $\Delta_{jk}$,
$\delta_{jk}$ ($j\neq k$):
\[
  e^{\Delta_{jk} + i\delta_{jk}} := 
  \frac{(\lambda_j-\lambda_k)(\lambda-\overline{\lambda}_k)}{(\lambda_j-\overline{\lambda}_k)(\lambda-\lambda_k)} \ .
\] \par \noindent
In the case $N=2$ we denote 
$\Delta:=\Delta_{12}\equiv \Delta_{21}$, $\delta_1:=\delta_{12}$, 
$\delta_2=\delta_{21}$ (compare \cite{Ci-twosol}).
The parameters are not independent. Indeed, 
$d_1(e^{\Delta-i\delta_1} - 1) \equiv d_2 (e^{\Delta+i\delta_2} -1)$.
Now we can write down a compact formula for ${\bf r}_{B+2}$:
\[ 
{\bf r}_{B+2} = {\bf r}_B + \frac{1}{2D}  
   \left( \begin{array}{c} d_1 \left( e^{Q_2} \cos\!\alpha_1^+ + 
	e^{-Q_2} \cos\!\alpha_1^- \right)    + 
   d_2 \left( e^{Q_1} \cos\!\alpha_2^+ + e^{-Q_1} \cos\!\alpha_2^- \right) \\
	d_1 \left( e^{Q_2} \sin\!\alpha_1^+ + 
	e^{-Q_2} \sin\!\alpha_1^- \right)    + 
   d_2 \left( e^{Q_1} \sin\!\alpha_2^+ + e^{-Q_1} \sin\!\alpha_2^- \right) \\ 
   d_+ \sinh\!(Q_1+Q_2) + d_- \sinh\!(Q_1-Q_2) +
	d_0 \sin\!(\alpha_1-\alpha_2) 
\end{array} \right)
\] \par \noindent	
where $D = \cosh\! Q_1 \cosh\! Q_2 \cosh\! \Delta + \sinh\! Q_1 \sinh\! Q_2 \sinh\! \Delta
   + \sinh\! \Delta \cos\!(\alpha_1-\alpha_2)$,
$\alpha_i^\pm = \alpha_i \pm \delta_i$, $d_\pm = (d_1 \pm d_2) e^{\pm\Delta}$
and $d_0 = d_1 \sin\!\delta_1 = - d_2 \sin\!\delta_2$.
The asymptotic behaviour of ${\bf r}_{2}$ can be calculated easily.
If $v^g_1 \neq v^g_2$ then we consider the limit $Q_2 \rightarrow \pm\infty$
(assuming $\mbox{$\mid Q_1  \mid$} \ll \mbox{$\mid Q_2  \mid$}$). Thus
\[ 
{\bf r}_{2} \ \ \stackrel{Q_2 \rightarrow \pm\infty}{\longrightarrow} \ \
 = \left( \begin{array}{c}  0 \\ 0 \\ \pm d_2 - x -\omega'(\lambda) t  \end{array} \right) +
  \frac{d_1}{\cosh\!(Q_1\pm\Delta)}
 \left( \begin{array}{c} \cos\!(\alpha_1\pm\delta_1) \\ \sin\!(\alpha_1\pm\delta_1) \\ \sinh\!(Q_1\pm\Delta)  \end{array} \right)
\] \par \noindent	
The shape and the velocity of the soliton do not change during the interaction.
The only result of the interaction is the {\bf phase shift}. 
In fact we have two phase shifts: the shift $\Delta_{ph}$ along ${\bf e}_3$ axis and
the shift of the angular variable $\alpha_1$. The first one is much more important
and can be measured in experiments. It can be easily calculated:
\[
 \Delta_{ph} = 2 d_2 + \frac{\Delta}{{\rm Im} \lambda_1} \ .
\] \par \noindent
The multi-soliton solutions are parameterized by complex eigenvalues 
$\lambda_1,\ldots,\lambda_N$ (in the sequel we denote $\lambda_k = a_k + i b_k$).
It is important to express the 
solutions by a set of $2N$ parameters which admit a physical interpretation,
like $d_k$, $v^g_k$, $v^{ph}_k$, $\Omega_k$, $\Delta_{ph}$ etc. (compare
\cite{MHR}).
The number of the physical parameters is much greater than $2N$ and
one may expect a lot of constraints which can be checked experimentaly.
The definition \refEq{dk} suggests the following change of variables: 
\begin{equation} \label{cov}
c_k + i d_k:=\frac{1}{\overline{\lambda}_k - \lambda}=\frac{a_k-\lambda+ib_k}{(a_k-\lambda)^2+b_k^2}
\ ,\qquad \lambda_k-\lambda = \frac{1}{c_k-i d_k} = \frac{c_k + id_k}{c_k^2+d_k^2}\ .
\end{equation} \par \noindent
Thus the parameters $c_k, d_k$ are expressed by $a_k, b_k$ and {\it vice
versa}. In particular we have:
\[
e^{\Delta_{jk} + i\delta_{jk}} := 
\frac{(d_j - d_k) + i(c_j - c_k)}{(d_j+d_k) + i(c_j-c_k)} \ .
\] \par \noindent
The physical meaning of $d_k$ is clear while the interpretation of
$c_k$ is, in general, a non-trivial problem. However, in the case of the
Localized Induction Equation (the case 
$\omega(\lambda) = - 2 \alpha \lambda^2$) the interpretation
is quite simple. Namely, $c_k = v^g_k / \Omega_k$  
 which means that 
$c_k$ is the distance travelled by the wave envelope during one full circle
of the wave maximum \cite{Ci-twosol}. 

Let us consider the case
of Localized Induction Equation with axial flow \refEq{axial}, $\omega(\lambda) = 
- 2 \alpha \lambda^2 - 4 \beta \lambda^3$. We have $\omega(0)=\omega'(0)=0$ and
the formulas \refEq{phys} assume the form
\begin{equation} \label{vvom} \begin{array}{c}
v^g_k = 4 \beta (b_k^2 - 3 a_k^2) - \alpha a_k \ , \quad
v^{ph}_k = 2 \alpha a_k^{-1} (b_k^2 - a_k^2) + 4 \beta (3 b_k^2 - a_k^2) \ , \\[1ex]
\Omega_k = - 4 (a_k^2 + b_k^2)(\alpha + 4 a_k \beta) \ .
\end{array} \end{equation} \par \noindent
Using \refEq{vvom} and \refEq{cov} we can easily check that
\[
    c_k = \frac{2\alpha}{v^{ph}_k-v^g_k} - \frac{16\beta}{\Omega_k} \ ,
	 \qquad a_k = \frac{2(v^g_k-v^{ph}_k)}{\Omega_k} \ .
\] \par \noindent
Finally, let us try to parameterize the $N$-soliton solution by
the parameters $d_k$ and $v^g_k$ which are most convenient from the
experimental point of view \cite{MHR}. It is sufficient to express $c_k$ in
terms of $d_k$ and $v^g_k$. By \refEq{vvom} and \refEq{cov} we have:
\[
   (c_k^2+d_k^2)^2 v^g_k = 4\beta(d_k^2-3c_k^2)-\alpha c_k (c_k^2+d_k^2) \ .
\] \par \noindent
Therefore, $c_k$ is a root of the algebraic equation of the 4-th order
with coefficients parameterized by $d_k$, $v^g_k$, $\alpha$ and $\beta$.

{\bf Compact formulas for  $N$-soliton surfaces}
can be derived from \refEq{iter} for an arbitrary $N$
\cite{Ci-Nsol}:
\[
{\bf r}_{B+N} = {\bf r}_B + \frac{i}{2} \left( \sum_{k=1}^N \frac{{\rm Im}
\lambda_k}{\mbox{$\mid \lambda-\lambda_k  \mid$}^2} - \sum_{j=1}^N \sum_{k=1}^N B_{kj} P_{kj} \right) \ ,
\] \par \noindent  
where
$B:=A^{-1}$ and $A$ is $N\!\times\! N$ matrix with  coefficients
$A_{jk} := i (\xi_k \overline{\xi}_j +1)(\lambda-\lambda_j)(\lambda-
           \overline{\lambda}_k)(\lambda_j-\overline{\lambda}_k)^{-1}$,
$P_{kj}$ are $2\!\times\! 2$ matrices defined by
\[
P_{kj} :=  \left( \begin{array}{cc} \xi_k \overline{\xi}_j & \xi_k \\ \overline{\xi}_j & 1 \end{array} \right) \ ,
\] \par \noindent
and, finally, $\xi_k$ are defined by \refEq{xi}.
Note that the surface ${\bf r}_{B+N}$ is expressed solely in terms of the 
background wave function $\Phi_B$ and $2N$ complex parameters:
$\lambda_k$ and $\gamma_k:=p_{k1}/p_{k2}$. Obviously, one can use $c_k$ and $d_k$
instead of $\lambda_k$.

We complete this section with  few remarks on the general $N$-soliton case.
To compute $N$-soliton addition to the  surface 
${\bf r}:=\Phi^{-1} \Phi,_\lambda |_{\lambda=\lambda_0}$ let assume the Darboux matrix in 
a general form 
\[
   D = \sum_{k=0}^N C_k (\lambda-\lambda_0)^k \ .
\] \par \noindent
The matrices $C_k$ are computed from the following linear system:
\[
    \sum_{k=0}^N C_k (\lambda_\nu - \lambda_0)^k \Phi(\lambda_\nu) p_\nu = 0 \ ,
    \quad (\nu=1,\ldots,Nn) \ ,
\] \par \noindent
where $\lambda_\nu \in {\bf C}$ and $p_\nu \in {\bf C}^n$ are constant.
Of course, one should take care of reductions which can result in
some constraints on $\lambda_\nu$, $p_\nu$.
The formula \refEq{r} assumes the form: 
$\tilde{\bf r} = {\bf r} +  \Phi^{-1} C_0^{-1} C_1 \Phi \ .$

\subsection*{Spectral problems from geometry}
\label{Lie}

It would be very important to be able to discern integrable classes of
surfaces. Here we present shortly how to approach this problem in a 
natural way using Lie symmetries.
Gauss-Weingarten equations, especially in the form \refEq{GW-matrix}, have very
similar form to the Zakharov-Shabat linear problem \refEq{ZS}. The only difference
is an absence of a spectral parameter.
It is well known that in many cases the spectral parameter is a group
parameter, i.e. there exists a symmetry of GMC equations (usually a simple one,
like a scaling or Lorentz or Galilean boost) which changes GW equations
\cite{Lu,Sa}.
The transformed GW equations contain explicitly the group parameter.
We have developed a systematic approach to study the problem
\cite{Ci-Gall,Ci-group,CGS-Heis,LST}. It consists in computing two algebras of
Lie symmetries: the algebra $A$ of symmetries of GMC equations and the algebra
$A'$ of symmetries of GW equations. Always $A' \subset A$. Reasonable
candidates for the spectral parameter are provided by vector fields $v$
such that $v\in A$ and $v\notin A'$. For more details and references see
\cite{Ci-Gall,Ci-FG,CGS-Heis}, compare also \cite{KV}.

Let us consider the following problem. We start from a given class of surfaces
(i.e., GW equations are given). Suppose that $A' \neq A$. Thus using an 
appropriate symmetry of GMC eqs.\ we can insert
a parameter into GW equations \refEq{GW-matrix} to obtain a linear problem of the
form \refEq{ZS}. Then we apply the Sym-Tafel formula \refEq{Sym}. 

Is the obtained class
of surfaces identical with the class we started from\,? We have no general
answer yet. In the following examples the answer is positive.

\begin{itemize}
\item {\bf Pseudospherical surfaces.} The fundamental forms 
    \refEq{ff-pseudospherical} with $\zeta=1$  and
   the symmetry $\tilde{u}=\zeta^{-1}u$, $\tilde{v}=\zeta v$ of the sine-Gordon equation
	yield the linear problem \refEq{problin-pseudospherical}.
\item {\bf Constant mean curvature surfaces.} The fundamental forms 
 \refEq{ff-Hconst} with $\kappa=0$, the symmetry  $\tilde{u}=u\,\cos\!\kappa - v\,\sin\!\kappa$ and
 $\tilde{v}=v\,\cos\!\kappa - u\,\sin\!\kappa$ of the equation \refEq{SHG}, and the gauge
transformation $\tilde{\Phi}=\exp(\kappa\,{\bf e}_2) \Phi$ yield the linear problem
\refEq{problin-Hconst}.
\item {\bf Spherical surfaces.}  The fundamental forms  \refEq{ff-spherical}
with $\kappa=0$  and the symmetry $\tilde{u}=u\,\cos\!\kappa - v\,\sin\!\kappa$ and
 $\tilde{v}=v\,\cos\!\kappa - u\,\sin\!\kappa$ of the elliptic sinh-Gordon equation \refEq{SHG}
  yield exactly the linear problem \refEq{problin-spherical}.
\item {\bf Bianchi surfaces.} The fundamental forms \refEq{ff-Bianchi-standard}
and the symmetry $\tilde{a} = a/\zeta$, $\tilde{b} = b\zeta$, $\tilde{f}= f/(1-2\lambda f)$ and
$\tilde{g}=g/(1+2\lambda g)$, where $\zeta=(1-2\lambda f)^{1/2} (1+2\lambda g)^{-1/2} = (1-2\lambda
\tilde{g})^{1/2} (1+2\lambda \tilde{f})^{-1/2}$, yield the linear problem
\refEq{problin-Bianchi}.
\end{itemize}	

\noindent In the second case it was necessary to perform a gauge
transformation dependent on the spectral parameter (otherwise,
starting from $H={\rm const}$ surfaces, one obtains spherical surfaces) \cite{DS}.

\section{Integrable geometries and Clifford algebras}
\label{CLIFF}

The examples presented above are associated with the ${\bf SU}(2)$
group. Recently, the soliton surfaces approach has been applied in more 
complicated cases (although the word ``surfaces'' is slightly misleading:
one can consider  submanifolds of higher 
dimensions). 
It is convenient to use Clifford algebras ${\cal C}(p,q)$ generated by elements
${\bf e}_1,\ldots,{\bf e}_m$ ($m=p+q$) satisfying 
\begin{equation}  \label{Cliff}
    {\bf e}_\mu {\bf e}_\nu + {\bf e}_\nu {\bf e}_\mu = 2 \eta_{\mu\nu} \ ,
\end{equation} \par \noindent
where $\eta_{\mu\nu}=0$ for $\mu\neq \nu$, $\eta_{\mu\mu}=1$ for 
$\mu=1,\ldots,p$ and $\eta_{\mu\mu}=-1$ for $\mu=p+1,\ldots,m$.
In the sequel the matrices $U_j$ of the linear problem are linear combinations
of ${\bf e}_\mu {\bf e}_\nu$. Then the function $\Psi$ assumes values in the group
Spin$(p,q)$ which is the double covering of ${\bf SO}(p,q)$. Note that 
${\bf SU}(2)$ can be identified with Spin$(3)$.

{\bf Isothermic surfaces} (isothermic immersions in ${\bf E}^3$)
can be defined as surfaces admitting {\em infinitesimal} isometries
preserving the mean curvature. Their curvature lines parameterized in a proper
way  form  a conformal coordinate system.  In other words, there exist local
coordinates $x^1,x^2$ in which fundamental forms read as follows \cite{Bi-izot}:
\begin{equation}   \label{I-II}
\begin{array}{c}
 I  =  e^{2\vartheta}( (dx^1)^2 + (dx^2)^2 )\ , \\[1ex]
 II  =  e^{2\vartheta}(k_2 (dx^1)^2 + k_1 (dx^2)^2)\ ,  
\end{array}
\end{equation} \par \noindent
\noindent where $k_1, k_2$ and $\vartheta$ depend on $x^1, x^2$.
Recently we found the following linear problem \cite{Ci-izot,CGS-izot}
which enable one to study isothermic surfaces using powerfull tools
of the theory of solitons:
\begin{equation}       \label{SPproblem}
\begin{array}{c}
      \Phi,_1 =  \frac{1}{2}\,{\bf e}_1 \left( -  \vartheta,_2 {\bf e}_2  - k_2 e^\vartheta {\bf e}_3 + 
        \lambda \sinh\!\vartheta\ {\bf e}_4  + \lambda \cosh\!\vartheta\ {\bf e}_5 \right)  \Phi \ ,    \\[1ex]
     \Phi,_2 = \frac{1}{2} \, {\bf e}_2 \left( -  \, \vartheta,_1 {\bf e}_1  - k_1 e^\vartheta {\bf e}_3
      + \lambda \cosh\!\vartheta {\bf e}_4      +      \lambda \sinh\!\vartheta\ {\bf e}_5 \right)  \Phi \ ,       
\end{array} \end{equation} \par \noindent
where ${\bf e}_k$  satisfy \refEq{Cliff} with $(\eta_{\mu\nu})={\rm diag}(1,1,1,1,-1)$
and $i{\bf e}_1{\bf e}_2{\bf e}_3{\bf e}_4{\bf e}_5=1$.
The Sym-Tafel formula needs some modification
to be applied to isothermic surfaces.
Namely, the formula \refEq{Sym}  for $\lambda=0$ defines 
a surface in 6-dim.\ space. Projecting the surface onto appropriate 
orthogonal 3-dim.\ subspaces we obtain a pair of dual isothermic surfaces.
Indeed, one can prove (see \cite{Ci-izot}) that
\begin{equation}                                                         \label{Sym'}  
   {\bf r}_{\pm} := \frac{1}{2} (1\mp {\bf e}_4 {\bf e}_5)\, \Phi^{-1}\Phi,_{\lambda}\, |_{\lambda =0}    
\end{equation} \par \noindent
are isothermic surfaces  immersed in Euclidean spaces
spanned by ${\bf e}_k({\bf e}_4\pm {\bf e}_5)$ ($k=1,2,3$), respectively.
The fundamental forms of the surface ${\bf r}_+$ are given exactly  by  \refEq{I-II}
while the ``dual surface'' ${\bf r}_-$ is the so called Christoffel transform of
${\bf r}_+$ \cite{Bi-izot,BP}.
The Darboux matrix for the linear problem \refEq{SPproblem} reads
\[
  D = \frac{{\bf e}_2}{\sqrt{\lambda^2 + \kappa_1^2}} \left( \kappa_1 (p_1 {\bf e}_1 + p_2 {\bf e}_2 
    + p_3 {\bf e}_3 ) + \lambda (\cosh\!\chi\ {\bf e}_4 + \sinh\!\chi\ {\bf e}_5) \right) \ ,
\] \par \noindent 
where $\kappa_1$ is a real parameter and $\chi,p_1,p_2,p_3$ 
are real functions which can be explicitly expressed by $\Phi$ evaluated
at $\lambda=-i\kappa_1$ \cite{Ci-dbt,Ci-izot}.
The corresponding Darboux-Bianchi transformation for the surfaces \refEq{Sym'},
\[
 \tilde{\bf r}_{\pm} = {\bf r}_{\pm} + \frac{2}{\kappa_1} e^{\pm\chi} \left( \pm p_1 e^{\mp\vartheta}
 {\bf r}_{\pm},_1 + p_2 e^{\mp\vartheta} {\bf r}_{\pm},_2 - p_3 {\bf n}_{\pm} \right) \ ,
\] \par \noindent
is identical with the classical Darboux-Bianchi transformation
for isothermic surfaces (compare \cite{Bi-izot}). 
The connection of the old branch of the classical differential geometry 
with the modern soliton theory {\it via} the linear problem \refEq{SPproblem}
has already started a series of new interesting developments in this field
\cite{BP,BHPP,CGS-izot}.

{\bf Space forms of dimension $n$ in ${\bf R}^{2n-1}$.}
The Lobachevsky plane  can not be immersed globally
in ${\bf R}^3$ and only local immersions (pseudospherical surfaces in ${\bf R}^3$) are
possible. Analogical situation has place for immersions of $n$-space forms
(spaces of constant curvature) for $n>2$. There are theorems on non-existence
of global immersions (see \cite{FP} and references cited therein) but  
local immersions are known (at least implicitly) \cite{Am77,Mo}. If the
curvature is constant and negative then there exists a coordinate system such
that all fundamental forms are diagonal. Moreover, the immersions turn out to
be parameterized by an orthogonal $n\!\times\! n$ matrix function $(a_{ij})$ (see, for
example,
\cite{ABT,Am77,TT}). The fundamental forms read:
\begin{equation} \label{forms}
  I = \sum_{j=1}^{n} a_{1j}^2 (dx^j)^2 \ , \qquad
  II^m = \sum_{j=1}^{n} a_{1j} a_{mj} (dx^j)^2  \ .
\end{equation} \par \noindent
where $m=2,\ldots,n$ and $\sum_{i=1}^{n} a_{ji} a_{jk} = \delta_{jk}$.
The coefficients $a_{ij}$ have to satisfy the
Gauss-Mainardi-Codazzi-Ricci equations. 
The equations are integrable: the B\"acklund transformation
\cite{TT}, the inverse scattering method \cite{ABT} and the loop group
approach \cite{FP} have been applied.  The following linear system with the
spectral parameter $\lambda$ is a modification of the ${\bf so}(n,n)$ spectral problem
presented by Ablowitz, Beals and Tenenblat (\cite{ABT}, see also
\cite{FP,Gol}):
\begin{equation} \label{nproblin}
\Psi,_j = \frac{1}{2} \left( \frac{\lambda^2-1}{2\lambda} a_{1j} {\bf e}_{1} + 
  \frac{\lambda^2+1}{2\lambda} \sum_{k=2}^n  a_{kj}{\bf e}_{k} 
  + \sum_{k=1}^n \gamma_{kj} {\bf e}_{n+k} \right) {\bf e}_{n+j} \Psi \ ,
\end{equation} \par \noindent
where $(a_{ij})$ and $(\gamma_{ij})$ are $n\!\times\! n$ matrices, $\gamma_{kk} = 0$ and
${\bf e}_{\mu}$ satisfy \refEq{Cliff} with $\eta_{\mu\nu} = \delta_{\mu\nu}$. 
The compatiblity conditions for the linear system \refEq{nproblin} 
yield that $(a_{ij})$ is orthogonal and
\begin{equation} \label{ccc} \begin{array}{c}
   a_{lj},_k = \gamma_{kj} a_{lk} \ , \quad 
	  	  \gamma_{ij},_k = \gamma_{ik} \gamma_{kj}  \ , \quad
  \gamma_{kj},_k + \gamma_{jk},_j + \sum_{i=1}^{n} \gamma_{ij} \gamma_{ik} = a_{1j} a_{1k} \ ,
\end{array} \end{equation} \par \noindent
where indices $i,j,k$ are distinct and, like the index $l$, run from $1$ to
$n$.  Let us define a map $F$ by the Sym formula
\begin{equation} \label{Sym1}
     F:= \Psi^{-1} \Psi,_\lambda |_{\lambda=1} \ .
\end{equation} \par \noindent
It defines (at least locally) an $n$-dimensional manifold (with possible
singularities) immersed, obviously, in the space of dimension
$n(2n-1)$ isomorphic with the Lie algebra ${\bf so}(2n)$.
One can show that in fact $F$ is confined to some 
Euclidean space of dimension $2n-1$.
Indeed, the unit tangent vectors are given by
$E_j = \frac{1}{2} \Psi^{-1} {\bf e}_1 {\bf e}_{n+j} \Psi$ while 
the normal space is spanned by
$ N_j := \Psi^{-1} {\bf e}_1 {\bf e}_j \Psi$ $(j=2,\ldots,n)$.
Differentiating $N_j$ we obtain elements of the tangent space,
$N_k,_j = a_{kj} E_k$.
The corresponding fundamental forms are given exactly by \refEq{forms}.
Thus we conclude that  the formula \refEq{Sym1} is an
expilcit expression for the $n$-dimensional manifolds of constant sectional 
curvature locally immersed in 
a Euclidean space of dimension $(2n-1)$ \cite{Ci-space}.

{\bf Lam\'e equations.}
We proceed to the general description of 
orthogonal coordinate systems in Euclidean spaces. The metric 
\begin{equation} \label{metric}
 I = H_1^2 (dx^1)^2 + \ldots + H_n^2 (dx^n)^2 \ ,
 \end{equation} \par \noindent
is induced by an orthogonal system in ${\bf E}^n$ provided that $H_k$
satisfy the Lam\'e equations \cite{Eis}
\begin{equation} \label{Lame} \begin{array}{c}
\left( \frac{H_j,_k}{H_k} \right),_k + \left( \frac{H_k,_j}{H_j} \right),_j + 
\sum_{i=1}^{n} \frac{H_j,_i H_k,_i}{H_i^2} = 0 \ , \\[3ex]
H_j,_{ik} = \frac{H_j,_k H_k,_i}{H_k} + \frac{H_i,_k H_j,_i}{H_i} \ .
\end{array} \end{equation} \par \noindent
Consider the following linear problem:
\begin{equation} \label{lproblin}
  \Psi,_j = {\bf e}_j(\lambda {\bf a}_j + {\bf b}_j) \Psi \ ,
\end{equation} \par \noindent
where ${\bf e}_1,{\bf e}_2\ldots,{\bf e}_{2n}$ generate the Clifford algebra ${\cal C}(2n)$,
${\bf a}_j := \frac{1}{2} \sum_{i=1}^{n} \alpha_{ji}{\bf e}_{n+i}$,  
${\bf b}_j := \frac{1}{2} \sum_{i=1}^{n} \beta_{ji}{\bf e}_i$ and $\beta_{jj}=0$.
The compatibility conditions for \refEq{lproblin} imply that the matrix 
$(\alpha_{ij})$ is orthogonal, the coefficients $\beta_{ij}$ are
given by  
$\beta_{jk} = - \alpha_{ji},_k/\alpha_{ki}$ and, finally, 
$H_k :=\alpha_{kj}$ satisfy (for  any fixed $j$) the Lam\'e equations \refEq{Lame}.
Let us define 
\[
    F:=\Psi^{-1} \Psi,_\lambda |_{\lambda=0} \ .
\] \par \noindent
The tangent vectors read 
$F,_j = \Psi^{-1} U_j,_\lambda \Psi |_{\lambda=0} = \Psi_0^{-1} {\bf e}_j {\bf a}_j \Psi_0$,
where $\Psi_0 := \Psi(x^1,\ldots,x^n;0)$ is contained in the subalgebra
generated by ${\bf e}_1,\ldots,{\bf e}_n$.
It is convenient to consider  projections $\Pi^k$ defined by
\begin{equation}
X = \sum_{i,j=1}^{n} X_{ji} {\bf e}_j {\bf e}_{n+i}  \longrightarrow 
\Pi^k X \equiv X^{(k)} := \sum_{j=1}^{n} X_{jk} {\bf e}_j \ ,
\end{equation} \par \noindent
and to apply them to $F$. As a result we obtain $F^{(1)}, F^{(2)},
\ldots, F^{(n)}$. Differentiating
$ F,_j^{(k)} = \alpha_{jk} \Psi_0^{-1} {\bf e}_j \Psi_0$ 
we derive that $\scalarproduct{F,_i^{(k)}}{F,_j^{(k)}} = \alpha_{ik} \alpha_{jk} \delta_{ij}$.
Therefore, each  map $F^{(k)}$ has the diagonal metric tensor \refEq{metric}.
Computing second derivatives we can check that $F^{(k)}$ defines an immersion
${\bf E}^n \rightarrow {\bf E}^n$, i.e., coordinates in ${\bf E}^n$ 
\cite{Ci-cross}. It would
be very interesting to interpret  the results of
\cite{KS} in our formalism.
By the way, the linear problem \refEq{SPproblem} can also be derived from
\refEq{lproblin} as a consequence of the compatibility conditions.
We just have to assume ${\bf a}_j$ as a linear combination of ${\bf e}_4$, ${\bf e}_5$
and ${\bf b}_j$ as a combination of ${\bf e}_1, {\bf e}_2, {\bf e}_3$.

\section{Conclusions}

The application of soliton techniques to the differential geometry 
revealed deep relations between these two areas 
\cite{Bob-tori,Bob,SR,S-0,Sym,Te}. In fact, many ideas 
of the soliton theory can be traced back to XIX century.
I would like to point out that the classical differential geometry 
studied a lot of special immersions and interesting transformations between
them \cite{Bi,Eis,Vek}. It is intriguing that the Sym-Tafel formula usually reconstructs 
immersions corresponding to a given GMC system. Then, using standard
methods of constructing soliton solutions, one is able to derive
the classical transformations of surfaces.

The other important application of the Sym-Tafel formula is the construction of
discrete surfaces. There is still not so clear how to define discrete analogues
of integrable classes of surfaces. Surprisinly effective way is provided by
applying the formula \refEq{Sym} to the corresponding discretization of the
linear problem (compare \cite{BP,BP-pseudo,Ci-space}).

The recent results of Doliwa and Santini  suggest even more
general approach. They consider immersions in a sphere $S^n$ and the
radius of the sphere is related to spectral parameter. In the limiting case
(infinite radius) $S^n \rightarrow {\bf R}^n$ and one can derive the Sym-Tafel
formula.  The results concerning integrable evolutions of curves are very
promising \cite{DS1,DS2}.

{\bf Acknowledgments.}
I would like to express my sincere thanks to  Antoni Sym for
many years of fruitful cooperation and to 
Decio Levi for many discussions and for his hospitality
during my several visits in Rome in the framework of the Rome-Warsaw
Universities agreement. 
I benefited greatly from discussions with Adam Doliwa. 
Thanks are due also to Reinhard Meinel, Gernot
Neugebauer and Heinz Steudel for helpful comments.
I am grateful to Joseph Krasil'shchik for interesting discussion during
the conference and for pointing me out a relevant part of \cite{KV}.
The work supported partially by the Polish Committee
of Scientific Researches (KBN grant 2 P03B 185 09).

\end{document}